# Ferromagnetism above 200 K in organic-ion intercalated CrSBr

Sofia Ferreira-Teixeira[1†], Daniel Tezze[1,2†], Maria Ramos[1,3], Covadonga Álvarez-García[1,2], Bertuğ Bayındır[4,5], Junhyeon Jo[1], Beatriz Martín-García[1,6], Maider Ormaza[2], Fèlix Casanova[1,6], Samuel Mañas-Valero[7], Eugenio Coronado[7], Hasan Sahin[4], Luis E. Hueso[1,6], Marco Gobbi[3,6*]

[1]CIC nanoGUNE BRTA, 20018 Donostia – San Sebastian, Basque Country, Spain

[2]Departamento de Polímeros y Materiales Avanzados: Física, Química y Tecnología, Facultad de Químicas (UPV/EHU), Apartado 1072, 2008, Donostia - San Sebastian, Basque Country, Spain

[3]Centro de Física de Materiales (CFM-MPC) Centro Mixto CSIC-UPV/EHU, 20018 Donostia - San Sebastian, Basque Country, Spain

[4] Department of Photonics, Izmir Institute of Technology, Izmir, Turkey

[5] Department of Physics, Izmir Institute of Technology, Izmir, Turkey

[6] IKERBASQUE, Basque Foundation for Science, 48009 Bilbao, Basque Country, Spain

[7] Instituto de Ciencia Molecular (ICMol), Universitat de València, 46980 Paterna, Spain

[†]These authors contributed equally to this work.
*Corresponding authors: marco.gobbi@ehu.eus

## Abstract

CrSBr is a van der Waals magnetic semiconductor exhibiting antiferromagnetic order below 140 K. It has emerged as a promising platform for engineering 2D magnetism because its intertwined electronic, optical, and magnetic properties can be profoundly modified via external stimuli such as electrical gating or magnetic fields. However, other strategies for tuning magnetism in layered materials, such as molecular intercalation, remain largely unexplored for CrSBr. Here, we demonstrate that the intercalation of tetramethylammonium (TMA) and tetrapropylammonium (TPA) ions into CrSBr induces a transition from antiferromagnetic to ferromagnetic order, while significantly enhancing the magnetic transition temperature to 190 K (TMA) and 230 K (TPA). The resulting intercalates are air-stable and exhibit large, hysteretic magnetoresistance exceeding 60% at 50 K in the TPA case. Besides, intercalation introduces symmetry-breaking structural changes in each CrSBr plane, revealed by Raman microscopy and corroborated by density functional theory (DFT) calculations. These findings highlight molecular intercalation as a powerful and versatile route to tailor the magnetic properties of CrSBr and unlock its potential to fabricate robust, high-temperature 2D magnetic devices.

## Keywords

Van der Waals magnets; CrSBr; 2D magnets; Molecular intercalation; Ferromagnetism; Magnetotransport; Air-stability.

## Introduction

Two-dimensional (2D) magnetic semiconductors offer an exciting platform to explore the interplay between spin, charge, and lattice degrees of freedom at the ultimate thin limit[1-2]. Among these materials, the air-stable van der Waals (vdW) compound CrSBr has recently emerged as a particularly intriguing system due to its unique combination of magnetic order, optical activity, and gate-tunable transport properties[2-3]. Its orthorhombic crystal structure hosts an A-type antiferromagnetic (AFM) ground state, in which Cr spins align ferromagnetically within each layer and couple antiferromagnetically between layers[4-6]. This AFM order sets in below a Néel temperature of approximately 140 K[7], while intralayer ferromagnetic (FM) correlations persist up to 180 K[6].

CrSBr features a direct bandgap, that gives rise to strong room-temperature photoluminescence (PL)[4], along with gate-tunable (magneto)transport. These properties are highly anisotropic[8-9]: the most intense PL is observed for excitation polarized along the crystallographic *b*-axis[10], which coincides with the magnetic easy axis and the most conductive direction[11-12].

A particularly appealing feature of CrSBr is the tunability of its physical properties via external stimuli. For example, electrical gating modulates carrier density and magnetoconductance[8, 11, 13], while applied magnetic fields induce a metamagnetic transition from AFM to FM interlayer alignment[14]. Beyond external fields, another powerful strategy to tailor the physical properties of vdW materials is intercalation, the insertion of atomic or molecular species between vdW layers[15-19]. This process often involves significant charge transfer and, in the case of bulky molecules, an expansion of the interlayer spacing[20-21]. These structural and electronic modifications can substantially influence the magnetic properties of the host[20, 22-29].

Despite the rapid rise of CrSBr in vdW magnetism, its intercalation chemistry remains largely unexplored. Initial studies reported that electrochemical intercalation of tetrabutylammonium ions enables the exfoliation of bulk crystals into thinner flakes[30]. These exfoliated flakes exhibit a transition from the pristine AFM to a FM state, while retaining a similar transition temperature[30]. Additionally, lithium intercalation has been shown to induce FM ordering, this time with an enhanced transition temperature of 200 K, although the resulting compound suffer from significant air sensitivity[31]. Thus, whether intercalation of molecular cations can effectively and reliably tune the magnetic behavior of CrSBr, while preserving its environmental stability, remains an open question.

Here, we show that the intercalation of two molecular cations, tetramethylammonium (TMA) and tetrapropylammonium (TPA), into CrSBr induces a transition from AFM to FM ordering. Remarkably, the resulting compounds are air-stable and exhibit a substantial enhancement of the transition temperature ($T_C$), which reaches 190 K for TMA-CrSBr and 230 K for TPA-CrSBr.

Both materials exhibit large and hysteretic magnetoresistance, exceeding 60% at 50 K for TPA-CrSBr. Raman spectroscopy reveals significant vibrational changes upon intercalation, consistent with a symmetry-lowering structural rearrangement. These findings are supported by density functional theory (DFT) calculations, which offer microscopic insight into the structural and electronic modifications induced by molecular intercalation. Our results establish molecular intercalation as an effective strategy to tune magnetism in CrSBr while preserving its environmental stability.

## Results and Analysis

TPA and TMA cations are intercalated into CrSBr via a galvanic method[32], as illustrated in Figure 1a. This approach exploits the spontaneous redox reaction between a magnesium (Mg) electrode and the CrSBr host, which act as the anode and cathode of a galvanic cell, respectively (Figure 1b). The method is compatible with both bulk crystals and exfoliated flakes. For bulk intercalation, a CrSBr crystal is clamped to a Pt plate, whereas in the case of flakes, CrSBr is exfoliated via micromechanical cleavage and transferred onto a conductive substrate. The CrSBr crystal or flakes are then electrically connected to a Mg strip through an external wire, and both are immersed in a solution of TMABr or TPABr. Under these conditions, Mg spontaneously oxidizes, injecting electrons into CrSBr and thereby reducing it. To preserve charge neutrality, molecular cations are inserted in the vdW gap of CrSBr. This galvanic process results in substantial electron doping, with each intercalated cation balanced by one injected electron.

The success of the intercalation process is confirmed through X-ray diffraction (XRD) analysis, shown in Figure 1c, which compares the diffraction patterns of pristine CrSBr, TMA-CrSBr, and TPA-CrSBr bulk crystals. Upon intercalation, the (00l) diffraction peaks systematically shift to lower angles, indicating an expansion of the interlayer spacing as the molecular cations are inserted between the vdW layers. Notably, no diffraction peaks associated with pristine CrSBr are observed in the intercalated samples, indicating that intercalation proceeds to completion. Using Bragg's law, we calculated the interlayer distances to be 7.95 Å for pristine CrSBr, 12.6 Å for TMA- CrSBr and 13.5 Å for TPA- CrSBr.

To investigate the effects of molecular intercalation on the magnetic properties of CrSBr, we performed vibrating sample magnetometry (VSM) measurements. Figure 1d displays the temperature dependence of the magnetization M(T) for bulk crystals of pristine, TMA-, and TPA-intercalated CrSBr, measured under an applied magnetic field of 0.1 T along the *b*-axis. As expected, the pristine CrSBr sample exhibits the characteristic antiferromagnetic (AFM) behavior, with a Néel temperature ($T_N$) of approximately 140 K[4, 6]. Upon intercalation with TMA and TPA, the magnetic order undergoes a striking transformation, as both intercalated samples display a clear FM response. The $T_C$ of the intercalated samples are significantly enhanced, reaching approximately 190 K for TMA-CrSBr and 230 K for TPA-CrSBr. These values were extracted from the temperature derivative of the M(T) curves (Supporting Information, Figure S1).

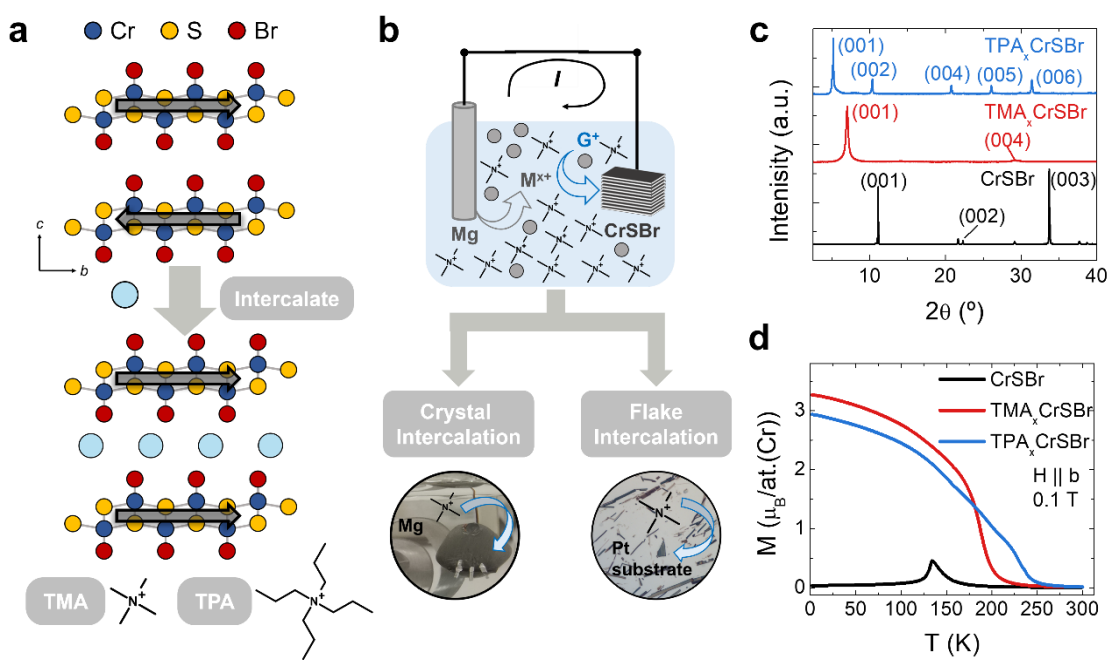

*Figure 1 Molecular intercalation of CrSBr. **a** Crystal structure of CrSBr before and after intercalation of organic cations (schematically shown as blue circles). The chemical structures of tetramethylammonium (TMA) and tetrapropylammonium (TPA) ions are also shown. The black arrows indicate the magnetic moment of each layer and its alignment regarding neighboring layers. **b** Schematic of the galvanic intercalation cell. CrSBr is electrically connected to a Mg electrode, and both are immersed in a solution of target guest molecules. Spontaneous oxidation of Mg leads to the intercalation of the molecular cations into CrSBr. **c** X-ray diffraction patterns of bulk crystals of pristine (black), TMA- (red) and TPA- (blue) intercalated CrSBr. **d** Temperature-dependent magnetization measured along the b-axis under an applied field of 0.1 T for pristine*

*CrSBr (black), TMA-CrSBr (red), and TPA-CrSBr (blue). Intercalation modifies the magnetic ordering temperature and overall moment.*

This enhancement in transition temperature is consistent with the recent reports of increased $T_C$ in Li-intercalated CrSBr[31]. Notably, the dependence of $T_C$ on the intercalated molecule suggests that organic-ion intercalation offers a versatile route to finely tailor the magnetic properties of CrSBr, going beyond the fixed doping level provided by Li.

Further insight into the nature of the ferromagnetic ordering in the intercalated samples is obtained from the M(T) curves. A Curie-Weiss analysis yields a positive Curie–Weiss temperature $\theta_{CW}$, larger than the actual $T_C$ for both TMA-CrSBr and TPA-CrSBr (Supporting Information, Figure S1). The positive values of $\theta_{CW}$ confirm the dominance of FM interactions, while the discrepancy with respect to $T_C$ suggests that the transition is not purely mean field. This interpretation is further supported by a critical exponent analysis (Supporting Information, Figure S1). Specifically, for TMA-CrSBr, we extracted a critical exponent β = 0.267 ± 0.002, which is lower than the mean-field value of 0.5 and closer to those expected for low-dimensional systems. This data suggests that molecular intercalation weakens the interlayer magnetic coupling and drives the system toward a quasi-2D magnetic regime. Considering that monolayer CrSBr is known to be FM, this analysis also implies that intercalation effectively imparts monolayer behavior to the bulk crystal, with each CrSBr layer acting as magnetically decoupled from its neighbors.

To address the effects of intercalation on magnetic anisotropy, we measured the magnetization as a function of the applied field along the three principal crystallographic axes (a, b, and c) of pristine and intercalated bulk crystals. These measurements, performed at 2 K, are presented in Figures 2a, b and c for pristine CrSBr, TMA-CrSBr and TPA-CrSBr, respectively. Pristine CrSBr is characterized by a low-field AFM state below the characteristic spin-flip transition at 0.25 T, followed by the magnetization saturation at 0.5 T, in agreement with previous reports[4]. No hysteresis is detected, confirming the absence of FM behavior in the pristine state.

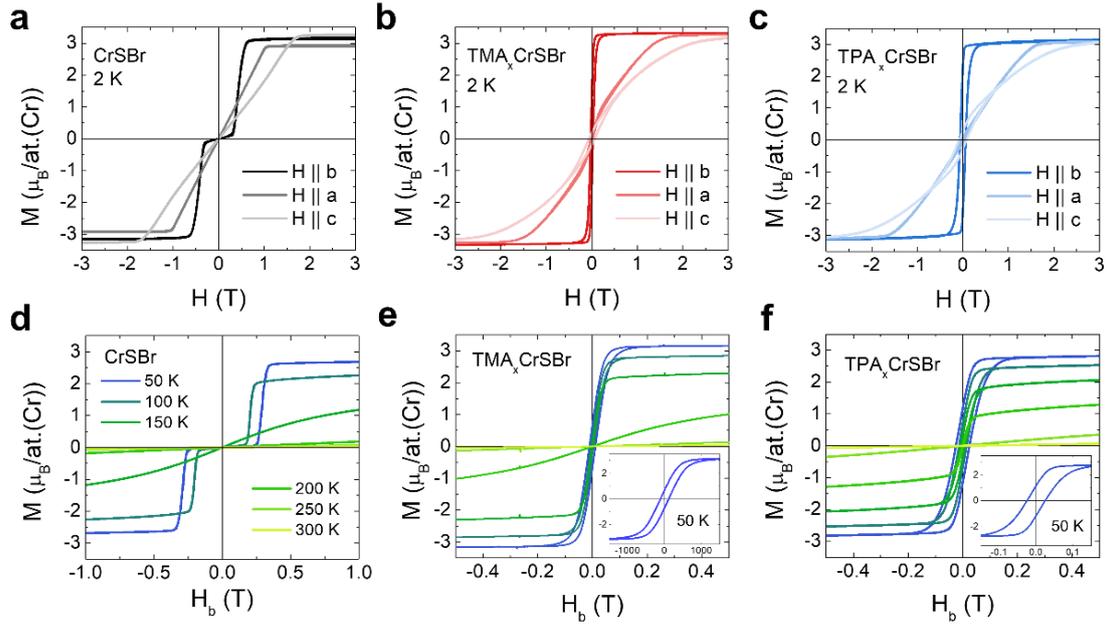

*Figure 2 Magnetic characterization of intercalated CrSBr. a b and c Magnetization as a function of external magnetic field aligned along the three crystallographic axes of CrSBr at 2 K for pristine CrSBr (a), TMA-CrSBr (b), and TPA-CrSBr (c). d, e and f Magnetization as a function of an external magnetic field along the b-axis of CrSBr at different temperatures for pristine CrSBr (a), TMA-CrSBr (b), and TPA-CrSBr (c). Insets in (e,f) highlight the magnetic hysteresis at 50 K.*

Strikingly, and consistently with the M(T) data, clear FM hysteresis loops are observed for the intercalated samples. Moreover, the triaxial anisotropy of pristine CrSBr is retained in the intercalated samples, with the b-axis remaining the easy-axis of magnetization, while the a-axis and c-axis continue to be the intermediate and hard axes, respectively. The coercive fields at 2 K are 25 mT and 60 mT for TMA-CrSBr and TPA-CrSBr, indicating that not only the $T_C$, but also the magnetic hysteresis is strongly influenced by the intercalated molecule. Notably, the magnetic properties of TMA-CrSBr remain unaltered in one year (Supporting Information, Figure S2).

Figures 2d, 2e, and 2f display the magnetization as a function of applied field along the b-axis measured at different temperatures for pristine, TMA-CrSBr, and TPA-CrSBr, respectively. In the intercalated samples, FM hysteresis loops persist at all temperatures up to their respective transition temperatures, demonstrating the stabilization of FM order upon intercalation. The insets highlight a zoomed-in view of the hysteresis loops at 50 K, showing that the coercive field is higher for TPA-CrSBr than for TMA-CrSBr, as for the measurement at 2 K. Above $T_C$, both intercalated samples exhibit a paramagnetic response, characterized by a linear dependence of magnetization on the applied field, without saturation.

Following the structural and magnetic characterization, we now explore the electronic transport properties of pristine and intercalated CrSBr. Figure 3a presents the resistivity of CrSBr, TMA-CrSBr, and TPA-CrSBr (see Methods), measured using a two-probe configuration with a voltage applied along the a-axis. The temperature dependence of the resistivity of pristine CrSBr follows the characteristic trend reported in the literature[4]. At high temperatures, it exhibits semiconducting behavior, with a thermally activated conduction mechanism. Below 230 K, the resistance increases but more smoothly, indicating a change in the conduction regime, while below 100 K, it rises more sharply again. In contrast, the intercalated samples do not exhibit distinct regime changes but instead show a smooth increase in resistance as the temperature decreases from 250 K down to 50 K.

To gain deeper insight into the conduction mechanisms, we analyzed the resistance data at low and high temperatures. At low temperature, the data for both pristine and intercalated samples are well fitted with an Efros-Shklovskii Variable Range Hopping (VRH) model for both pristine and intercalated samples, consistent with the previous reports[13, 31, 33] (Supporting Information, Figure S3). At higher temperatures, the data are well described by a thermally activated conduction mechanism:

$$R = R_0 \exp\left(\frac{E_a}{k_B T}\right) \qquad (2)$$

where $E_a$ is the activation energy. This behavior was previously reported for pristine and Li-intercalated CrSBr[31, 33]. In our case, we extracted activation energies of 218 meV for pristine CrSBr, which is consistent with prior reports[13], 114 meV for TMA-CrSBr and 65 meV for TPA-CrSBr (Supporting Information, Figure S3). The decrease in activation energy upon intercalation suggests that doping effects modify the density of states of the conducting electrons. However, the $E_a$ values extracted here are higher than the 37 meV reported for Li-intercalated CrSBr[31], indicating that the doping level achieved via molecular intercalation is lower than that induced by lithium. Moreover, the temperature ranges over which the thermally activated conduction dominates differ between samples. For pristine CrSBr the range is between 230 K and room temperature, whereas for the intercalates it extends from 210 K (TMA-CrSBr) or 240 K (TPA-CrSBr) up to room temperature. These temperature regimes align well with the paramagnetic regions observed in the magnetization measurements of the intercalated crystals, suggesting a coupling between the magnetic and electronic properties in intercalated CrSBr.

To investigate the interplay between magnetism and charge transport, we performed magnetoresistance (MR) measurements on TMA-CrSBr and TPA-CrSBr at 100 K, with the field applied along b and c (Figure 3b and c), and the current along a. Here we define MR as:

$$MR = \frac{R(H) - R(1\ \text{T})}{R\ (1\ \text{T})} \tag{3}$$

where R(H) is the resistance at field H, and R(1 T) is the resistance at 1 T.

The MR recorded for both samples is strongly influenced by the relative spin alignment between magnetic layers. At high magnetic fields (~ 1T), the magnetization within each layer aligns with the field direction, yielding a parallel spin configuration characterized by low resistance (Figure 3b and c). As the field is swept through the coercive region, the weakly coupled magnetic layers reverse their magnetization at slightly different field values, leading to an intermediate state with misaligned spins, characterized by a higher resistance.

This behavior is reminiscent of the mechanism underlying giant magnetoresistance (GMR) in magnetic multilayers[34], where the resistance depends on the relative orientation of FM layers separated by a nonmagnetic spacer even when the current flows in plane. Intercalated CrSBr effectively behaves as a self-assembled GMR multilayer, in which conductive magnetic layers are naturally spaced by nonmagnetic organic molecules. However, unlike conventional GMR stacks that switch between fully parallel and antiparallel magnetization states, intercalated CrSBr transitions only into a misaligned configuration, without reaching a fully antiparallel state, which is a unique venue for realizing organic spintronics[35-36]. To reflect this analogy, we define MR by normalizing the resistance to its value at 1 T, corresponding to the parallel state, as common in GMR studies[37].

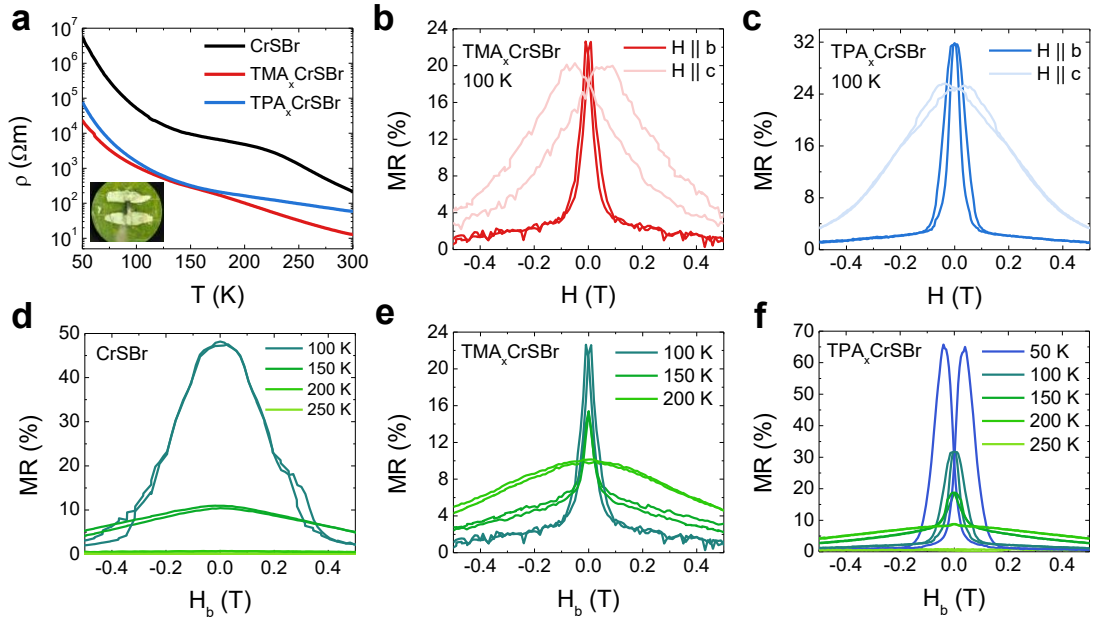

*Figure 3 Electrical transport characterization of intercalated CrSBr. a Temperature dependence of the resistivity for pristine CrSBr (black), TMA-CrSBr (red), and TPA-CrSBr (blue). The measurement was performed on bulk crystals (inset) in a two-probe configuration, applying 1 V along the a-axis of CrSBr. The inset shows an optical photograph of a measured TMA crystal. b and c Magnetoresistance (MR) recorded with an external magnetic field aligned along the b and c crystallographic axes of CrSBr at 100 K for TMA- CrSBr (b) and TPA-CrSBr (c). d, e and f MR recorded with an external magnetic field aligned along b-axis of CrSBr at different temperatures for pristine CrSBr (d), TMA-CrSBr (e) and TPA-CrSBr (f).*

Figures 3d, 3e, and 3f display the temperature dependence of MR for pristine, TMA-CrSBr, and TPA-CrSBr. For pristine CrSBr, the MR follows a quadratic field dependence with no hysteresis for temperatures above the Néel temperature, consistent with prior literature[4]. Below the transition temperature, high values of MR, approximately 47%, are obtained at 100 K, with no hysteresis, in accordance with previous results[4]. In contrast, for the intercalated crystals, the MR behavior closely follows the magnetization as a function of field as discussed previously, reaching up to 65% for TPA-CrSBr at 50 K.

We also intercalated few-nm-thick exfoliated CrSBr flakes (Figure 4). Figure 4a shows optical images of a CrSBr flake on a Pt-coated Si/SiO$_2$ substrate before and after intercalation. A significant color change is observed, with the intercalated flake appearing much darker. This suggests modifications in its electronic structure, likely due to charge transfer effects that alter optical absorption and reflectivity.

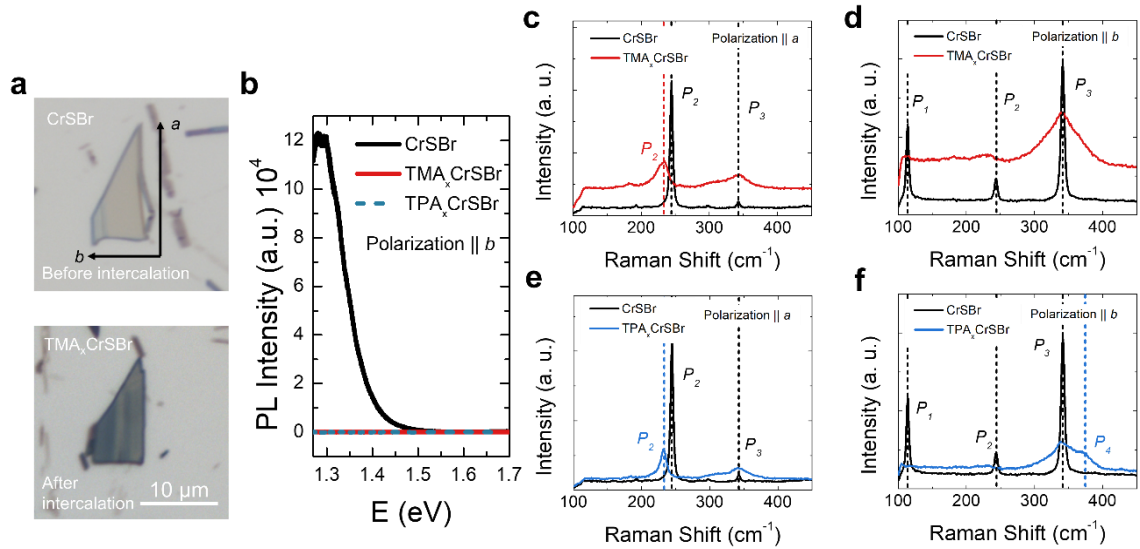

*Figure 4 Raman and Photoluminescence of intercalated CrSBr. a Optical Image of a CrSBr flake before (top) and after (bottom) TMA intercalation. b Photoluminescence of flakes of pristine CrSBr (black), TMA-CrSBr (red) and TPA-CrSBr (dashed-blue) at room temperature. No photoluminescence is observed for the intercalated flakes. c and d Raman spectra of pristine CrSBr (black) and TMA-CrSBr flake (red) with excitation light polarized along the a and b-axes of CrSBr, respectively. e and f Raman spectra of pristine CrSBr (black) and TMA-CrSBr flake (blue) with excitation light polarized along the a and b-axis of CrSBr, respectively.*

Figure 4b shows the PL spectra of pristine and intercalated CrSBr flakes with polarization along the b-axis. The pristine CrSBr flake exhibits a PL signal centered around 1.3 eV, in agreement with previous reports[38]. In contrast, for both TMA- and TPA-intercalated flakes, the PL signal completely vanishes across the entire measured energy range. This suggests that the intercalated materials no longer behave as semiconductors, likely due to a heavily doped state.

Micro-Raman spectra of pristine and intercalated flakes are shown in Figures 4c–f, for excitation light polarized along the a- and b-axes. The Raman spectrum of pristine CrSBr displays three characteristic peaks at 114 cm$^{-1}$ (P1), 244 cm$^{-1}$ (P2) and 342 cm$^{-1}$ (P3), consistent with previous studies[38]. These peaks also exhibit the reported polarization dependence[38]. Specifically, when the excitation is polarized along the a-axis, P1 is suppressed while P2 becomes more intense. Upon intercalation of TMA and TPA, notable changes are observed. P1 is suppressed in both polarization configurations, while P2 shifts from 244 cm$^{-1}$ to 233 cm$^{-1}$. P3 retains its same position but broadens significantly, particularly for b-axis polarization. In the case of TPA intercalation, an additional shoulder emerges near 374 cm$^{-1}$ under b-axis polarization.

To understand the structural and electronic modifications induced by intercalation, we also performed density functional theory (DFT) calculations on bilayer CrSBr in both its pristine and TMA-intercalated form (see Methods). Total energy minimization calculations reveal that TMA molecules approaching preferentially adsorb near the surface of the CrSBr layers, displacing four adjacent Br atoms from their equilibrium positions (Figure 5a and Supporting information, Note 1). This atomic rearrangement, driven by electron donation from TMA, breaks the original crystal symmetry and leads to changes in the Raman response.

To capture these effects quantitatively, we computed the Raman activity by evaluating the changes in the macroscopic polarizability tensor (see Supporting Information, Note 1). The calculated Raman spectrum for pristine CrSBr and TMA-CrSBr are shown in Figure 5b. The calculated Raman spectrum for pristine CrSBr shows good qualitative agreement with our experimental measurements. Additionally, our calculations indicate that upon intercalation, the three prominent peaks observed in the pristine material undergo significant changes. Specifically, the structural relaxation and increased interatomic distances induce phonon softening, which accounts for the experimentally observed shift of P2.

For P3, associated with out-of-plane vibrations of Cr and S atoms, the calculations predict a lifting of degeneracy and the appearance of a low-intensity mode corresponding to in-plane vibrations (see Supporting Information, Figure S5). Experimentally, P3 broadens considerably, especially for b-axis polarization. While the individual modes cannot be resolved, likely due to structural disorder and thermal broadening at finite temperature, the broadening is consistent with the predicted coexistence of out-of-plane and in-plane vibrational contributions.

Beyond accounting for the changes in the vibrational properties, DFT calculations also shed light on the AFM-to-FM transition due to intercalation. Each TMA molecule donates its delocalized electron to the adjacent CrSBr layers, effectively doping the material and altering its electronic and magnetic states (see Supporting Information, Note 1, Figure S4).

The electronic density of states (DOS) calculated for pristine CrSBr bilayer reveals the material is a semiconductor composed of FM layers coupled antiferromagnetically, resulting in a spin-degenerate DOS. This is consistent with the known properties of pristine CrSBr[11]. Upon intercalation, the DOS profile of the structure changes dramatically (Figure 5d). As a result of conduction band filling due to electrons transferred from TMA, the interlayer magnetic interaction switches from AFM to FM, breaking spin degeneracy and producing a net spin polarization. These theoretical predictions are in line with the experimental observation of a transition to a ferromagnetic ground state upon intercalation.

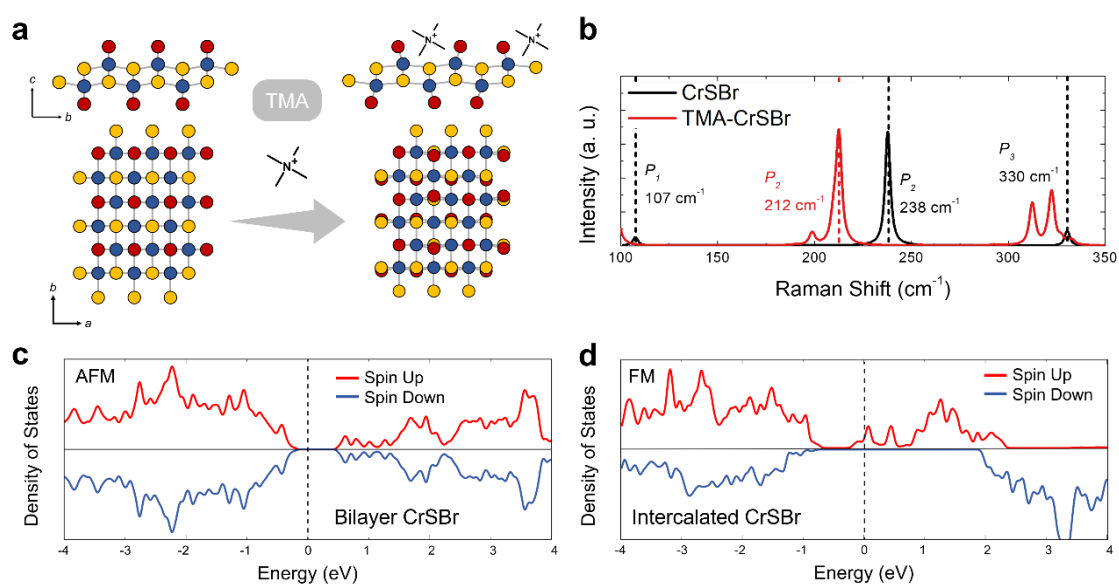

*Figure 5 Theoretical calculation of electronic and structural properties of intercalated CrSBr. a Structural change of the intercalated CrSBr. The interaction between TMA molecules and the CrSBr surface modifies the atomic positions of the Br atoms, shifting their alignment with the Cr and S atoms. b Theoretically calculated Raman modes for bilayer CrSBr (black) and TMA-Intercalated CrSBr (red). c Spin resolved Density of States of bilayer pristine CrSBr. The DOS is the same for both spin up (red) and down (blue), confirming the AFM ground state. d Spin resolved Density of States of bilayer CrSBr intercalated with the TMA molecules. The DOS is different for*

*spin up and down, showing that the ground state of intercalated CrSBr is FM. The Fermi level is set to zero in both DOS plots.*

## Conclusions

In summary, molecular intercalation presents a powerful route to engineer the properties of CrSBr. This method enables tunability of the material's structural, magnetic, and electronic characteristics, offering a new pathway to design air stable vdW materials with tailored functionalities. Our data indicates that intercalated CrSBr is composed of magnetically decoupled FM layers separated by non-magnetic spacers. The observed intercalation-induced transition from an AFM to a FM state, along with modifications in electronic transport properties, opens opportunities for applications in spintronic devices and magnetoelectronics. Future studies could explore the intercalation of different molecular species to further expand the range of tunable properties in CrSBr and the insertion of the intercalated CrSBr in multifunctional devices.

## Methods

<u>Crystal Growth:</u> CrSBr crystals were synthesized by Chemical Vapor Transport and characterized by powder and crystal X-ray diffraction, energy dispersive X-ray analysis, high-resolution transmission electron microscopy, superconducting quantum interference device magnetometry and temperature-dependent single crystal diffraction, as reported previously[9].

<u>Galvanic Intercalation:</u>

<u>X-ray diffractometry:</u> XRD measurements were carried out using an Empyrean diffractometer (PANalytical) on bulk crystals and on exfoliated flakes supported on Pt substrates. A copper cathode was used as X-ray source. Both the wavelengths $K_{\alpha 1}$ (1.5406 Å) and $K_{\alpha 2}$ (1.5443 Å) were employed to maximize the intensity of the diffracted beam.

<u>µ-Raman spectroscopy and photoluminescence spectroscopies:</u> A Renishaw® in-Via Qontor micro-Raman instrument was used to collect the Raman spectra of the samples at room temperature using a 532 nm laser as excitation source. A 50× focusing objective providing a laser spot size of ~1.5 µm was used during data acquisition. Raman measurements of the intercalated flakes were performed in a vacuum chamber (Linkam, HFS350EV-PB4) provided with an optical window for spectroscopy, reaching pressures down to $10^{-5}$ hPa. This setup was used because the flakes exhibited sensitivity to laser irradiation uniquely under ambient conditions. Polarization-dependent Raman measurements were performed by inserting a half-wave plate in the optical path, with its fast axis at 45º relative to the incident electric field, effectively rotating the light polarization by 90º. Photoluminescence was acquired using the same system, configured to detect over a broader spectral range. Due to the grating limitation of our Raman setup, photon energies below ~1.25 eV are not detectable, and thus the low-energy side of the PL peak of pristine CrSBr is not captured.

Magnetization measurements: Magnetization as a function of temperature or magnetic field measurements were carried out using a physical properties measurement system (PPMS, Quantum Design) in vibrating sample magnetometer (VSM) mode.

Electronic Transport measurements: Electrical transport measurements were performed in a physical property measurement system (PPMS, Quantum Design) with a rotational sample stage. The measurements were performed using two-probes by placing the bulk crystals on indium pads and connecting it to the pads and the PPMS puck with silver paste (RS Pro Silver Conductive Paste RS 186-3600). A Kethley 2636 was used to apply a DC voltage across the bulk crystals and measure the current. The resistivity of each crystal was calculated by measuring the width and length of the expected conduction channel using an optical microscope, and the thickness through the X-ray diffractometer. However, due to the non-uniformity of the crystals, which also does not allow us to know directly what is the effective conduction channels of the crystals, the measurements of these geometric factors give rise to high uncertainties, which we expect to be up to ~40%.

Theoretical calculations: Optimized atomic structures, electronic configurations, and magnetic ground states of TMA molecules, CrSBr layers, and their mutual interactions were investigated using first-principles calculations based on density functional theory (DFT), using the Vienna Ab initio Simulation Package (VASP)[39]. Exchange-correlation interactions were treated within the generalized gradient approximation (GGA) using the Perdew-Burke- Ernzerhof (PBE) functional[40]. To properly account for the localized nature of Cr 3d orbitals, the DFT+U method, as proposed by Dudarev[41], was applied with an effective on-site Coulomb interaction parameter of $U_{eff}$ = 3.0 eV. A kinetic energy cutoff of 500 eV was used for the plane-wave expansion. The total energy convergence criterion was set to $10^{-5}$ eV. To eliminate spurious interactions between periodic images in the out-of-plane direction, a vacuum spacing of 15 Å was introduced.

After determining the ground state of the monolayer crystal, simulation of multilayered structures was performed on bilayers. TMA intercalation calculations were carried out by inserting different percentages of TMA molecules between sufficiently large bilayer supercells. One electron transfer from each TMA molecule to CrSBr crystals and the local doping caused by this were obtained similarly for different adsorption percentages. Raman activity spectra were obtained using the finite-difference method to evaluate the change in the macroscopic polarizability tensor $\alpha_{ij}$ with respect to atomic displacements along each normal vibrational mode $Q_k$[42].

## Author Contributions

## Acknowledgments

The authors acknowledge technical support from Roger Llopis. This work was supported under Projects PID2021-128004NB-C21 and PID2021-122511OB-I00 funded by Spanish MCIN/AEI/10.13039/501100011033 and by ERDF A way of

making Europe; and under the María de Maeztu Units of Excellence Programme (Grant CEX2020-001038-M). This work was also supported by the FLAG-ERA grant MULTISPIN, via the Spanish MCIN/AEI with grant number PCI2021-122038-2A. S.F.-T. acknowledges financial support from the European Union via Marie Skłodowska-Curie grant agreement number 101106104 CHEERS. B.M.-G. and M.G. thank support from "Ramón y Cajal" Programme by the Spanish MICIU/AEI/10.13039/501100011033 and European Union NextGenerationEU/PRTR (grant nos. RYC2021-034836-I and RYC2021-031705-I, respectively).